\documentclass[aps,prl,twocolumn,groupedaddress,showpacs]{revtex4}
\usepackage{epsfig}

\def\q{{\bf q}}
\def\p{{\bf p}}
\def\k{{\bf k}}
\def\u{{\bf u}}
\def\v{{\bf v}}
\def\8{\infty}
\def\oh{\frac{1}{2}}

\def\undertext#1{\vtop{\hbox{#1}\kern 1pt \hrule}}

\def\VEV#1{\langle\,#1\,\rangle}

\def\be{\begin{equation}}
\def\ee{\end{equation}}
\def\bea{\begin{eqnarray} & &}
\def\eea{\end{eqnarray}}

\def\rf#1{(\ref{#1})}

\def\rf#1{(\ref{#1})}

\def\rfs#1{Eq.~\rf{#1}}

\begin{document}


\title{Quantum phase transitions across a $p$-wave Feshbach resonance}


\author{V. Gurarie}
\author{L. Radzihovsky}
\author{A. V. Andreev}
\affiliation{Department of Physics, University of Colorado,
Boulder CO 80309}


\date{\today}

\begin{abstract}
  We study a single-species polarized Fermi gas tuned across a narrow
  $p$-wave Feshbach resonance. We show that in the course of a
  BEC-BCS
  crossover the system can undergo a magnetic field-tuned quantum phase
  transition from a $p_x$-wave to a $p_x+i p_y$-wave superfluid. The
  latter state, that spontaneously breaks time-reversal symmetry,
  furthermore undergoes a topological $p_x+ i p_y$ to $p_x+ i p_y$
  transition at zero chemical potential $\mu$.  In two-dimensions,
  for $\mu>0$ it is characterized by a Pfaffian ground state
  exhibiting topological order and non-Abelian excitations familiar
  from fractional quantum Hall systems.
\end{abstract}

\maketitle

Recent progress in controlling atomic interactions via an s-wave
Feshbach resonance (FR) has led to a realization of an s-wave paired
fermionic superfluid (SF) in a variety of degenerate atomic
gases\cite{Jin2004,Grimm2004,Ketterle2004}.  By tuning through FR such
paired superfluid has been studied along the crossover between the BCS
regime of weakly-paired, strongly overlapping Cooper pairs, and the
BEC regime of tightly bound, weakly-interacting diatomic molecules.

Although the study of this crossover has received considerable
attention\cite{Timmermans2001,Holland2001},
its equilibrium properties are already {\it qualitatively } well
described by early seminal works\cite{Leggett1980,Nozieres1985}.
Furthermore, for a narrow FR, the thermodynamics can even be
accurately computed analytically across the full range of the BCS-BEC
crossover, with corrections controlled by the ratio of the FR width to
Fermi energy\cite{UsUnpublished}.

In contrast, superfluids paired at a {\em finite} angular-momentum
are characterized by richer order parameters (as exemplified by a
$p$-wave paired $^3$He, heavy-fermion compounds, and $d$-wave
high-T$_c$ superconductors), and therefore admit phase transitions
within the SF phase.  Recent observation of a $p$-wave
FR\cite{Ticknor2004,Schunck2004} in $^{40} K$ and $^6 Li$,
prepared in identical spin states, demonstrates a promising
mechanism for a realization of $p$-wave superfluidity,  attracting
recent theoretical attention \cite{Ho2004,Ohashi2005,Botelho2004}.
Because Pauli exclusion principle prevents identical fermionic
atoms from scattering via $s$-wave channel, such $p$-wave
scattering dominates.  The tuning of the atomic interaction
through such FR should allow access to aforementioned phase
transitions and provides strong motivation for our study.

\begin{figure}[bth]
\centering \setlength{\unitlength}{1mm}
\begin{picture}(60,62)(0,0)
\put(-12,-3){\begin{picture}(70,70)(0,0)
\includegraphics{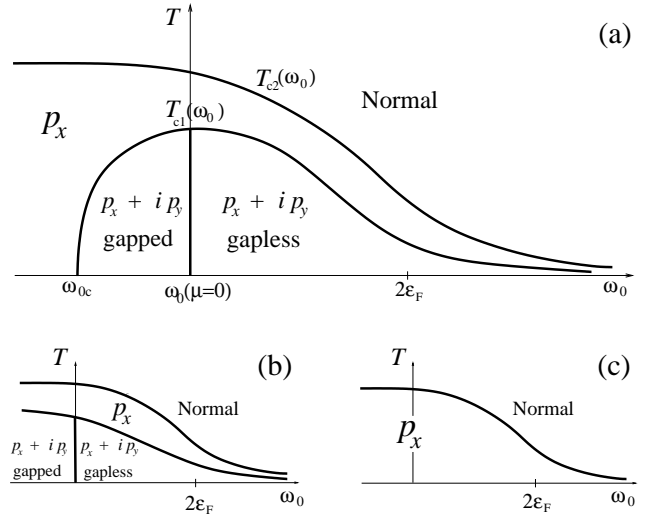}
\end{picture}}
\end{picture}
\caption{(a) A schematic temperature-detuning phase diagram near a
$p$-wave FR, for intermediate splitting of $m=0$, $\pm 1$ FR (see
below). It shows a 2nd-order $p_x$ to $p_x+i p_y$ SF transition at
$\omega_{0c}$ as well as a topological phase transition at
$\omega_{0}(\mu=0)$ between two different $p_x+i p_y$ phases. (b),
(c) show alternative phase diagrams for smaller and larger values
of FR splitting, respectively.
} \label{phasediagramFig}
\end{figure}

In this Letter, we study a $p$-wave superfluidity in a
single-species polarized Fermi gas tuned across a $p$-wave FR. We
show analytically that at low $T$, in the course of a BEC-BCS
crossover this system can exhibit a second-order quantum phase
transition as a function of FR detuning $\omega_0$ (energy of the
quasi-bound molecular state) from a $p_x$-wave to a time-reversal
breaking $p_x+i p_y$-wave superfluid. In three dimensions (3D),
for a positive chemical potential $\mu>0$, the $p_x+i p_y$-wave
superfluid exhibits gapless single particle excitations with
momentum ${\bf p} = \pm p_F\,\hat{\bf
  z}$, that become gapped for $\mu<0$. Consequently, on quite general
grounds we predict that in addition to above $p_x$ to $p_x+i p_y$
phase transition, the latter SF must furthermore undergo a
topological quantum phase transition at $\omega_{0}(\mu=0)$, from
a gapped to gapless $p_x+i p_y$ SF state\cite{Volovik2004}
(Fig.~\ref{phasediagramFig}(a,b)) Associated thermodynamic changes
(e.g., a heat capacity changing from activated to power-law form)
should be observable.

Moreover, when confined to 2D, for $\mu>0$, the $p_x+i p_y$ superfluid
is described by a fully-gapped Pfaffian (Moore-Read) ground state
\cite{MR}, that exhibits a subtle topological order characterized by
excitations with non-Abelian statistics, familiar from fractional
quantum Hall systems. At $\omega_{0}(\mu=0)$, this fully-gapped
superfluid has been demonstrated to undergo a topological quantum
phase transition to another strongly-coupled gapped $p_x+i p_y$-wave
superfluid \cite{VolovikBook,Read2000}.

The accuracy of our theory is controlled by the ratio of the width of
the FR (molecular decay rate) to $\epsilon_F$.  Because the low-energy
decay rate of a finite angular momentum diatomic molecule is
suppressed by the centrifugal barrier, our theory can be made
arbitrarily accurate by reducing the atom density, $n$.

We start with a Hamiltonian of spinless fermionic atoms, resonantly
interacting through a diatomic molecular state with internal angular
momentum (``spin'') $\ell=1$:
\begin{eqnarray}
\label{eq:ham} H &=& \sum_{p} \epsilon_p ~\hat a^\dagger_{\bf p}
\hat a_{\bf p} + \sum_{{\bf p}, \alpha} \left(\epsilon_\alpha +
{\epsilon_p \over 2} \right)
\hat b_{\alpha{\bf p}}^\dagger \hat b_{\alpha {\bf p}}\\
&+&\sum_{{\bf p},{\bf q},\alpha} ~{g_p \over \sqrt{V}} ~p_\alpha
\left( \hat b_{\alpha {\bf q}} ~ \hat a^\dagger_{\p+{\q\over 2}}
~\hat a^\dagger_{-\p+{\q \over 2}} + h. c. \right).\nonumber
\end{eqnarray}
Here $\hat a_{\bf p}^\dagger$ is the creation operator of an atom with
momentum ${\p}$ and kinetic energy $\epsilon_p=p^2/2m$. $\hat
b^\dagger_{\alpha{\q}}$ creates a molecule with momentum ${\q}$,
kinetic energy $\epsilon_q/2$, and $\alpha$ labels its spin
polarization. $V$ is the volume of the system. $\epsilon_\alpha$
controls the molecular rest energy related to FR detuning, that can in
general be polarization dependent\cite{Ticknor2004}. The
fermion-molecule matrix element $g_p$ is a constant $g_0$ for $p \ll
\Lambda$, quickly dropping off to zero for $p \gg \Lambda$. The
ultraviolet (UV) cutoff $\Lambda$ is associated with the smallest
length scale in the problem, the size of the molecular
(closed-channel) bound state.

We first study elastic resonant scattering of two atoms with
momenta $\p$ and $-\p$ into states $\p'$ and $-\p'$. This will
enable us to relate the bare parameters $\epsilon_\alpha$
appearing in $H$ to the experimental FR detuning $\omega_\alpha$.
In vacuum this process can only proceed through an intermediate
molecular bound state, created by $\hat{b}^\dagger_{\alpha 0}$, at
zero momentum and energy $E=p^2/m$. The scattering amplitude
$f(\p,\p')$ is determined by the 4-fermion amputated correlation
function. It can be deduced from the molecular Green's function at
zero momentum,
$D_{\alpha \beta}(E) = \delta_{\alpha \beta} \left[
E-\epsilon_\alpha-\Pi(E) \right]^{-1}$,
expressed in terms of the molecular self-energy (fermionic
polarization ``bubble'') $\Pi(E)$, given by $\Pi(E)= \int {d
\omega \over 2 \pi} {d^3 k \over (2 \pi)^3} {{2 \over 3} i g^2_k
~k^2 \over \left(E-\omega-{k^2 \over 2m}+ i 0
\right)\left(\omega-{k^2 \over 2m} + i 0\right)}$.  Evaluating the
integrals (with $\epsilon_\Lambda \gg E$), we arrive at an {\em
exact} expression for the $p$-wave scattering amplitude
\begin{equation}
\label{eq:scatnonc}f(\p,\p')=\sum_{\alpha} {-(m g_0^2/2\pi)\,
p_\alpha p'_\alpha\over {p^2 \over m}\left(1+c_2 \right) -
\epsilon_\alpha+c_1 + i {m g_0^2  \over 6 \pi} p^3},
\end{equation}
with $c_1={m\over 3\pi^2}\int_0^\infty d k ~k^2 g^2_k$ and
$c_2={m^2\over 3\pi^2}\int_0^\infty d k ~g^2_k$ quasi-molecular
size (UV cutoff) dependent constants.

Firstly, we observe that for a rotationally-invariant case of
$\epsilon_\alpha=\epsilon_0$, \rfs{eq:scatnonc} reduces to the
$\ell=1$ ($p$-wave) case of the most general expression for the
scattering amplitude, $f_\ell(p)=[p^{-2\ell} F_\ell(p^2) - ip]^{-1}$,
for scattering from a central potential in angular momentum channel
$\ell$, obtained simply from unitarity and analyticity.  $F_\ell(p^2)$
is a real function Taylor-expandable in powers of its
argument\cite{LL}, that in our case can be read off from
\rfs{eq:scatnonc} and given {\em exactly} (for $\epsilon_\Lambda\gg
E$) by $F_1(p^2) = -A+B p^2,$ where $A=-\left(\epsilon_0-c_1\right){6
\pi \over m g_0^2}$ is the inverse scattering volume, and $B=-{6 \pi
\over m^2 g_0^2} \left(1+c_2 \right)$.

Hence, $H$ in \rfs{eq:ham} correctly reproduces the low-energy
two-body physics, regardless of the details of real interatomic
interactions. The physical detuning position of the FR (corresponding
to the molecular state that is real/virtual when it is
negative/positive) is given by the low-energy poles of the scattering
amplitude $f({\bf p},{\bf p'})$, $\omega_\alpha \equiv
{\epsilon_\alpha - c_1 \over 1+c_2}$, expressed in terms of the bare
Hamiltonian parameters $\epsilon_\alpha$. In contrast to the $s$-wave
case, where such renormalization only shifts the position of the FR,
here it also leads to a multiplicative renormalization.

We now proceed to the main topic of the Letter, namely to the
study of thermodynamics of a fermionic atomic gas at a finite
total density $n$, near a $p$-wave FR described by $H$,
\rfs{eq:ham}.  It is convenient to work in the grand-canonical
ensemble, $H_\mu = H - \mu \hat{n}$
, with the
chemical potential $\mu$ determined by the average total atom
density $n = {1 \over V} \sum_{\bf p} \left( \sum_{\alpha} 2 \VEV{
\hat b_{\alpha \bf p}^\dagger \hat b_{\alpha \bf p}} + \VEV{ \hat
a^\dagger_{\bf p} \hat a_{\bf p}} \right). $

$H_\mu$ cannot be diagonalized exactly. However, as we will show
below, at $T=0$ the system is in a paired SF state, characterized
by a finite $\VEV{\hat{b}_{\alpha 0}} = \sqrt{V} B_\alpha$ for
{\em any} value of detuning, $\omega_\alpha$. This can be easily
seen by noting that for a large positive detuning
$\omega_\alpha\gg 2\epsilon_F$, there is a large gap to creating
(quasi-)molecules, $\hat{b}_\alpha$. Therefore they can be
integrated out to lead to an effective BCS model, with an
attractive $p$-wave four-Fermi interaction of strength
$\lambda_{BCS}=g_0^2/(\omega_\alpha - 2\mu)$, with virtual
molecules mediating the attractive interactions between fermions.
The resulting finite Cooper-pair field
$\VEV{\hat{a}_{-\p}\hat{a}_{\p}}$ in this BCS regime then always
induces a finite (albeit exponentially small)
$\langle{\hat{b}_{\alpha 0}}\rangle$ through a resonant $g_0$
coupling. In the opposite BEC regime, $\omega_\alpha <
2\epsilon_F$, we will show that $\mu$ decreases, tracking closely
$\omega_\alpha/2$, and the energy can be lowered by binding a
fraction of atoms into Bose-condensed $p$-wave molecules,
stabilized by Pauli-blocking.

Hence, to make progress we approximate the molecular field $\hat
b_{\alpha\p}$ by its condensate value $\VEV{\hat{b}_{\alpha 0}} =
\sqrt{V} B_\alpha$, neglecting bosonic excitations. This reduces
$H_\mu$ to a quadratic, mean-field BCS form (with $B_\alpha$ playing
the role of the Hubbard-Stratonovich field, i.e., the gap function),
that can be easily diagonalized by a Bogoliubov transformation.

The key point is, that, independent of $\omega_\alpha$,
fluctuations about this molecular saddle-point approximation are
suppressed in powers of a dimensionless coupling \vspace{-0.2cm}
\begin{equation}
\gamma={\sqrt{2}\over 3\pi^2}g_0^2\epsilon_F^{1/2}m^{5/2},
\label{gamma}
\end{equation}
proportional to the ratio of the FR width to Fermi energy.  $\gamma$
can always be made sufficiently small by simply reducing the density
$n=(2m\epsilon_F)^{3/2}/(6\pi^2)$~\cite{swave}. Hence, for $\gamma\ll
1$, total number of excited molecular bosons $\sum_{\alpha, \q \not =
  0} \VEV{\hat b^\dagger_{\alpha \q} \hat b_{\alpha \q}}$ is small,
and above molecular mean-field approximation is {\em quantitatively}
accurate, with corrections $O(\gamma)$. In the BCS regime, this
reduces to the standard criterion for the validity of the BCS
mean-field corresponding to the gap much smaller than $\epsilon_F$,
namely Cooper pairs that are large compared to interatomic
spacing\cite{BCScutoff}.  However, in contrast to the standard BCS
model, the $\gamma \ll 1$ criterion is far less stringent and allows a
quantitatively accurate analysis even in the molecular BEC regime,
where from the BCS prospective the atoms are strongly coupled with
$\mu$ much smaller than the atomic gap. In this regime the limitation
of the molecular mean-field theory simply corresponds to a small
depletion of the molecular condensate.

Hence for $\gamma\ll 1$, we can accurately compute the free energy
$F(B_\alpha)$.
As usual, its minimization with respect to the complex order
parameter ${\bf B}$, $\partial F[{\bf B}]/\partial B_\alpha^*=0$
leads to a BEC-BCS gap equation,
\begin{equation}\hspace{-0.05cm}
\label{eq:gap} \left( \epsilon_\alpha - 2 \mu \right) B_\alpha =
 B_\beta \int {d^3 k \over (2 \pi)^3} {g^2_k k_\alpha k_\beta  \over E_k}
 \tanh \left(\frac{E_k}{2k_B T} \right),
\end{equation}
whose solution, together with the total atom conservation equation
\begin{equation}
\label{n_conserve}
\hspace{-0.11cm} n=2 n_b + n_a
\end{equation}
determines the molecular condensate amplitude, ${\bf B}$ and the
chemical potential, $\mu$ as a function $\omega_\alpha$, $T$ and $n$.
In above, $n_a(\mu)=\oh\int {d^3 k\over (2 \pi)^3} \left[ 1- {{k^2
      \over 2m} - \mu \over E_k }\tanh \left(\frac{E_k}{2k_B T}
  \right) \right]$ is the number of open-channel BCS-paired atoms,
with $E_k=\sqrt{\left({k^2 \over 2 m}-\mu \right)^2 + 4 g^2_k |{\bf
    B}\cdot\k|^2}$ the Bogoliubov atomic quasi-particle spectrum, and
$n_b=\sum_{\alpha {\bf q}} \VEV{\hat b^\dagger_{\alpha \q} \hat
  b_{\alpha \q}}$ the total number of closed-channel molecules.

The complex vector order parameter ${\bf B}$ encodes the three
molecular states $\psi_{m}$ with spin projection $m=0,\pm 1$
through the relations $\psi_0= B_z$, $\psi_1=-{1 \over \sqrt{2}}
\left(B_x+i B_y \right)$, $\psi_{-1}={1 \over \sqrt{2}}
\left(B_x-i B_y \right)$ \cite{LL}. It can be parameterized in
terms of two real vectors \be {\bf B}={\bf u} + i {\bf v}, \ee
that by a gauge transformation ${\bf B} \rightarrow e^{i \varphi}
{\bf B}$ can always be conveniently chosen to be orthogonal, with
${\bf u}\cdot{\bf v}=0$. Consequently, if either one (but not
both) of the vectors vanishes (e.g., ${\bf v}$), then the
projection of the angular momentum on the axis parallel to the
other vector (${\bf u}$) is equal to zero. If, on the other hand,
$u=v$ (with $u=|{\bf u}|$ and $v=|{\bf v}|$), then the projection
of the angular momentum on $\v \times \u$ is equal to
$1$~\cite{LL}.

Focusing first on the  $T=0$ case, by integrating \rfs{eq:gap} we
find the free energy in this $u,v$ representation
\begin{eqnarray}
\label{eq:free} &&\frac{F(\u,\v)}{1+c_2}= \sum_\alpha
\left(u_\alpha^2+v_\alpha^2 \right) \left[ \omega_\alpha - 2 \mu
+a_1 \ln\left\{ a_0 \left(u+v \right)\right\} \right]
\nonumber\\
&&+ a_1 {u^3+v^3\over u+v} + a_2\left[ \left(u^2+v^2 \right)^2 +
\oh \left( u^2-v^2 \right)^2 \right],
\end{eqnarray}
with $a_1=\left( {2\over 1+c_2}\gamma \sqrt{\mu
/\epsilon_F}\right)\mu$, $a_2=\left({8\over 5} {c_2 \over
1+c_2}\gamma \right) \epsilon_F/n$, $a_0= e^{5/6}
(\mu/\epsilon_F)^{1/4} (\gamma/8n)^{1/2} $.
As usual, the logarithmic term arises from the Fermi surface
($k^2/2m \sim \mu$) contribution to the integral of \rfs{eq:gap}.
It is only a good estimate for $\mu > 0$ when the logarithm itself
is large, and vanishes otherwise. On the other hand, the quartic
terms $a_2 \big[\left( {\bf B^* \cdot B} \right)^2 + \oh \left|
{\bf B}\cdot {\bf B} \right|^2 \big]$ in $F$ arise from momenta
$k\sim\Lambda$.

Minimization of $F(u,v)$ leads to the phase diagram as a function
of detuning $\omega_\alpha$, displayed in
Fig.\ref{phasediagramFig}.  For $\omega_0\gg 2 \epsilon_F$ (BCS
regime), the molecular density is vanishingly small and atom
conservation, \rfs{n_conserve}, constrains $\mu$ to be close, but
slightly  below $\epsilon_F$.  The minimum of $F(u,v)$ is then
controlled by the balance of the quadratic and logarithmic terms
in $u$ and $v$, with the quartic terms exponentially small. In the
{\em isotropic} case with molecular spin-independent FR,
$\omega_\alpha=\omega_0$, $F$ is invariant under simultaneous
rotation of ${\bf u}$ and ${\bf v}$, and exhibits three (plus
symmetry related) local extrema: (i) normal state with $u=v=0$,
(ii) $p_{x}$-wave superfluid with $u \neq 0$, $v=0$, and (iii)
$p_x+i p_y$-wave superfluid with $u=v \not = 0$. The absolute
minimum of $F$ is given by (iii) with $u=v\approx \frac{1}{2 a_0
e}~e^{-(\omega_0-2 \epsilon_F)/a_1}$, and the condensation energy
ratio ${\cal{R}}=F_{p_x+i p_y}/F_{p_x}\approx e/2 > 1$, robustly
giving $p_x+i p_y$ as the ground state in the BCS regime,
consistent with the $A_1$ superfluid ground state of a fully
spin-polarized $^3$He.\cite{Vollhardt,Anderson1961,Ho2004}.

In the complementary $\omega_0 < 2 \epsilon_F$ crossover regime,
it becomes favorable to pair-up a finite fraction of atoms into
closed-channel Bose-condensed molecules at energy $\omega_0$. This
forces $\mu$ to track $\omega_0/2$ below $\epsilon_F$.
Consequently, the condensate density $n_b \approx|B|^2$ is no
longer exponentially small, growing to a finite fraction of the
overall density $n$, and as a result the logarithmic term in
\rfs{eq:free} becomes negligible. The condensate density $|B|^2$
is then primarily determined by the atom number \rfs{n_conserve},
with the gap equation simply constraining $\mu$ to track
$\omega_0/2$ to the accuracy ${\cal O}(\gamma)$.
In the {\em isotropic} case, it is clear from \rfs{eq:free} that
although the quadratic ${\bf B^*} \cdot {\bf B}$ and the first quartic
$\left( {\bf B^*} \cdot {\bf B} \right)^2$ terms are invariant under
rotation in the $u-v$ space, the positive coefficient of the $\left|
  {\bf B} \cdot {\bf B} \right|^2 =(u^2-v^2)^2$ quartic term
guarantees that $F(u,v)$ is again minimized by
$u=v\approx{1\over2}\sqrt{n-n_a(\omega_0/2)}$, corresponding to the
time-reversal-breaking $p_x+i p_y$-wave superfluid ground state. In
this idealized isotropic case, the finite $T$ transition from the
normal to the $p_x+i p_y$ superfluid state is described by the complex
O$(3)$ vector model, $F=t {\bf B^* \cdot B} + \lambda_1 \left( {\bf
    B^* \cdot B} \right)^2 + \lambda_2 \left| {\bf B}\cdot {\bf B}
\right|^2$.

However, in the experimental realization, the magnetic dipole
interaction generically splits the FR into two $m=0$ and $m=\pm 1$
resonances\cite{Ticknor2004}. Choosing $x$-axis along the external
magnetic field, this can be captured in our model \rfs{eq:ham} by
taking $\omega_x =\omega_0-\delta\not = \omega_y=\omega_z \equiv
\omega_{0}$.  This orbital easy axis (experimentally $\delta>0$)
anisotropy substantially enriches the phase diagram arising from
minimization of $F[\u,\v]$, Eq.\ref{eq:free}. A technically
involved analysis\cite{UsUnpublished} predicts for an intermediate
FR splitting $1/(1+c_2) <\delta/(n a_2) < 5/(4c_2)$ (Fig.~1(a)),
that for $\omega_0<\omega_{0c}$ the free-energy is minimized by a
$\u=u\hat{x}$, $\v=0$ state corresponding to a $p_x$-wave $m=0$
superfluid. At low $T$, for  $\omega_0 > \omega_{0c}$, we instead
find that the free-energy is minimized by a $u\neq v\neq 0$,
characterizing a state that breaks time-reversal symmetry and is a
coherent superposition of  $p_x$ and $p_x + i p_y$ superfluids
(that we refer to as simply a $p_x + i p_y$ state).  It is easy to
see that the quantum phase transition from the $p_x$ to the $p_x+i
p_y$ superfluid is second-order. The global (gauge) U$(1)$
symmetry is spontaneously broken in the $p_x$ superfluid, and
rotational O$(3)$ symmetry is explicitly broken (down to O$(2)$)
by the dipole interaction that acts like Ising (easy axis)
anisotropy.  Hence it can be shown \cite{UsUnpublished} that the
$T=0$ $p_x$ to $p_x+i p_y$ transition at $\omega_{0c}$ is in the
4D O$(2)$ (XY) universality class. The corresponding critical
point controls the shape of the low $T$ phase boundary with
$T_{c1} (\omega_0) \sim \left| \omega_0 - \omega_{0c} \right|^{z
\nu}$, and $z=1$ and $\nu={1 \over 2}$.

From the general structure of $F$ near $T_{c2}$ ($\sim \epsilon_F$ for
$\omega_0 \lesssim 2 \epsilon_F$ and exponentially small in the BCS
regime), where the nonanalytic logarithmic term in \rfs{eq:free} is
suppressed (through the hyperbolic tangent in \rfs{eq:gap}), it is
possible to show that for an arbitrary small FR splitting $\delta>0$
the normal-superfluid transition is always into the $p_x$ state (with
the transition in the 3D XY universality class), in contrast to the
isotropic case, $\delta=0$, where it is always into the $p_x+i p_y$
superfluid\cite{UsUnpublished}.

For a sufficiently small FR splitting $\delta/(n a_2)<1/(1+c_2)$
(Fig. 1(b)), the $T=0$ phase transition disappears and at low $T$
the $p_x + i p_y$ superfluid extends to all detuning, undergoing a
transition to the $p_x$ state at $T_{c1}\left(\omega_0\right)$
($\sim \epsilon_F$ for $\omega_0 \lesssim 2 \epsilon_F$) only near
the normal-to-superfluid $T_{c2}\left(\omega_0 \right)$, below
which the $p_x$ superfluid is confined in a sliver of
$T_{c1}<T<T_{c2}$ set by $\ln \left[T_{c2}/T_{c1} \right] \sim
\delta$. This is consistent with the isotropic case $\delta=0$,
where $p_x+i p_y$ state extends all the way up to $T_{c2}$. In
contrast, for a large FR splitting (Fig. 1(c)), $\delta/(n
a_2)>5/(4 c_2)$, it is clear from $F$, \rfs{eq:free} that the
dipole anisotropy dominates all other terms and leads to the $p_x$
state over the full $T-\omega_0$ phase diagram. However, it is
important to note that for a very large FR splitting, $\delta$,
given long energy relaxation times, near the $m=1$ resonance the
atomic gas may remain in a metastable $p_x+i p_y$ state, despite
the existence of a lower energy $p_x$ state.

The analysis presented in this Letter establishes the stability of
$p_x+ip_y$ SF phase over a large portion of the phase diagram, in an
experimentally realizable system in which the chemical potential $\mu$
can be directly tuned through zero by adjusting FR detuning
$\omega_0$. This allows us to take advantage of the established
literature\cite{Leggett1980,Volovik2004,VolovikBook,Read2000}, that
demonstrates that such SF state undergoes a topological phase
transition at $\mu=0$ between two types of $p_x+i p_y$ SFs. In 3D
these are distinguished by the existence of gapless atomic excitations
for $\mu>0$ (due to nodal points at $\p=\pm p_F \, \hat {\bf z}$) and
a fully gapped spectrum for $\mu<0$. In 2D (to which our computation
of $F$ can be easily extended without any qualitative changes
\cite{UsUnpublished}) both of these states are fully gapped but
nevertheless are distinguished by topological
order\cite{VolovikBook,Read2000}. Thus our work explicitly shows that
degenerate Fermi gases near a $p$-wave FR give a concrete realization
of non-Abelian topological states, which are of interest in strongly
correlated systems with possible applications to fault-tolerant
quantum computing\cite{Kitaev2003}.  Detecting phases of $p$-wave
superfluids and transitions between them (illustrated in
Fig.~\ref{phasediagramFig}) in cold atomic gases remains an open
problem.

We acknowledge support by the NSF DMR-0321848 (L.R.), DMR-9984002
(A.A.), and the Packard Foundation (A.A., L.R.), and thank J.
Bohn,  and D. Jin for discussions.

{\sl Note added:} After this paper was posted, a manuscript by
Chi-Ho Cheng and S.-K. Yip \cite{Yip2005} appeared that studied
the same problem. For large $\big( \delta/(n a_2)>5/(4c_2) \big)$
and small $\big( \delta/(n a_2)<1/(1+c_2) \big)$ Feshbach
resonance splitting their results  coincide with ours (summarized
by our Figs.~1 (b,c)). However, results disagreed (see Fig.~1(a)
in version 2 of our manuscript) for the narrow range of the
intermediate values of the Feshbach resonance splitting,
$1/(1+c_2)<\delta/(n a_2)<5/(4c_2) $, for reasons at the time not
understood by either groups. Upon further analysis we found that
Cheng and Yip's predictions \cite{Yip2005} for this narrow range
of FR splitting was correct. Our mistake stemmed from not
appreciating the dominant (in the narrow intermediate range of FR
splitting) contribution to the atom number coming from the long
tail in the atomic momentum distribution function. Correction of
this mistake is reflected in the updated version of Fig.~1(a), but
does not affect any other result in our original manuscript. We
credit Cheng and Yip for the correct version of Fig.~1(a).
Detailed analysis will be presented in the manuscript currently in
preparation \cite{Gurarie2005}.

We regret that due to our oversight  this note did not make it
into the version of our manuscript published in Physical Review
Letters.


\bibliography{pwave}

\begin{thebibliography}{25}
\expandafter\ifx\csname natexlab\endcsname\relax\def\natexlab#1{#1}\fi
\expandafter\ifx\csname bibnamefont\endcsname\relax
  \def\bibnamefont#1{#1}\fi
\expandafter\ifx\csname bibfnamefont\endcsname\relax
  \def\bibfnamefont#1{#1}\fi
\expandafter\ifx\csname citenamefont\endcsname\relax
  \def\citenamefont#1{#1}\fi
\expandafter\ifx\csname url\endcsname\relax
  \def\url#1{\texttt{#1}}\fi
\expandafter\ifx\csname urlprefix\endcsname\relax\def\urlprefix{URL }\fi
\providecommand{\bibinfo}[2]{#2}
\providecommand{\eprint}[2][]{\url{#2}}

\bibitem[{\citenamefont{{Regal {\it et al.}}}(2004)}]{Jin2004}
\bibinfo{author}{\bibfnamefont{C.~A.} \bibnamefont{{Regal {\it et al.}}}},
  \bibinfo{journal}{Phys. Rev. Lett.} \textbf{\bibinfo{volume}{92}},
  \bibinfo{pages}{040403} (\bibinfo{year}{2004}).

\bibitem[{\citenamefont{{Bartenstein et al}}(2004)}]{Grimm2004}
\bibinfo{author}{\bibfnamefont{M.}~\bibnamefont{{Bartenstein et al}}},
  \bibinfo{journal}{Phys. Rev. Lett.} \textbf{\bibinfo{volume}{92}},
  \bibinfo{pages}{120401} (\bibinfo{year}{2004}).

\bibitem[{\citenamefont{{Zwierlein et al}}(2004)}]{Ketterle2004}
\bibinfo{author}{\bibfnamefont{M.~W.} \bibnamefont{{Zwierlein et al}}},
  \bibinfo{journal}{Phys. Rev. Lett.} \textbf{\bibinfo{volume}{92}},
  \bibinfo{pages}{120403} (\bibinfo{year}{2004}).

\bibitem[{\citenamefont{{Timmermans {\it et al.}}}(2001)}]{Timmermans2001}
\bibinfo{author}{\bibfnamefont{E.}~\bibnamefont{{Timmermans {\it et al.}}}},
  \bibinfo{journal}{Phys. Lett. A} \textbf{\bibinfo{volume}{285}},
  \bibinfo{pages}{228} (\bibinfo{year}{2001}).

\bibitem[{\citenamefont{{Holland {\it et al.}}}(2001)}]{Holland2001}
\bibinfo{author}{\bibfnamefont{M.}~\bibnamefont{{Holland {\it et al.}}}},
  \bibinfo{journal}{\prl} \textbf{\bibinfo{volume}{87}},
  \bibinfo{pages}{120406} (\bibinfo{year}{2001}).

\bibitem[{\citenamefont{Leggett}(1980)}]{Leggett1980}
\bibinfo{author}{\bibfnamefont{A.}~\bibnamefont{Leggett}}, in
  \emph{\bibinfo{booktitle}{Modern Trends in the Theory of Condensed Matter}}
  (\bibinfo{publisher}{Springer-Verlag}, \bibinfo{address}{Berlin},
  \bibinfo{year}{1980}), pp. \bibinfo{pages}{13--27}.

\bibitem[{\citenamefont{Nozi\`eres and Schmitt-Rink}(1985)}]{Nozieres1985}
\bibinfo{author}{\bibfnamefont{P.}~\bibnamefont{Nozi\`eres}} \bibnamefont{and}
  \bibinfo{author}{\bibfnamefont{S.}~\bibnamefont{Schmitt-Rink}},
  \bibinfo{journal}{J. Low Temp. Phys.} \textbf{\bibinfo{volume}{59}},
  \bibinfo{pages}{195} (\bibinfo{year}{1985}).

\bibitem[{\citenamefont{Gurarie et~al.}()\citenamefont{Gurarie, Radzihovsky,
  and Andreev}}]{UsUnpublished}
\bibinfo{author}{\bibfnamefont{V.}~\bibnamefont{Gurarie}},
  \bibinfo{author}{\bibfnamefont{L.}~\bibnamefont{Radzihovsky}},
  \bibnamefont{and} \bibinfo{author}{\bibfnamefont{A.~V.}
  \bibnamefont{Andreev}}, \bibinfo{note}{unpublished}.

\bibitem[{\citenamefont{{Ticknor {\it et al.}}}(2004)}]{Ticknor2004}
\bibinfo{author}{\bibfnamefont{C.}~\bibnamefont{{Ticknor {\it et al.}}}},
  \bibinfo{journal}{Phys. Rev. A} \textbf{\bibinfo{volume}{69}},
  \bibinfo{pages}{042712} (\bibinfo{year}{2004}).

\bibitem[{\citenamefont{{Schunck {\sl et al.}}}()}]{Schunck2004}
\bibinfo{author}{\bibfnamefont{C.~H.} \bibnamefont{{Schunck {\sl et al.}}}},
  \eprint{cond-mat/0407373}.

\bibitem[{\citenamefont{Ho and Diener}(2005)}]{Ho2004}
\bibinfo{author}{\bibfnamefont{T.-L.} \bibnamefont{Ho}} \bibnamefont{and}
  \bibinfo{author}{\bibfnamefont{R.~B.} \bibnamefont{Diener}},
  \bibinfo{journal}{Phys. Rev. Lett.} \textbf{\bibinfo{volume}{94}},
  \bibinfo{pages}{090402} (\bibinfo{year}{2005}).

\bibitem[{\citenamefont{Ohashi}(2005)}]{Ohashi2005}
\bibinfo{author}{\bibfnamefont{Y.}~\bibnamefont{Ohashi}},
  \bibinfo{journal}{Phys. Rev. Lett.} \textbf{\bibinfo{volume}{94}},
  \bibinfo{pages}{050403} (\bibinfo{year}{2005}).

\bibitem[{\citenamefont{Botelho and de~Melo}()}]{Botelho2004}
\bibinfo{author}{\bibfnamefont{S.}~\bibnamefont{Botelho}} \bibnamefont{and}
  \bibinfo{author}{\bibfnamefont{C.~A. R.~S.} \bibnamefont{de~Melo}},
  \eprint{cond-mat/0409357}.

\bibitem[{\citenamefont{Klinkhamer and Volovik}(2004)}]{Volovik2004}
\bibinfo{author}{\bibfnamefont{F.~R.} \bibnamefont{Klinkhamer}}
  \bibnamefont{and} \bibinfo{author}{\bibfnamefont{G.~E.}
  \bibnamefont{Volovik}}, \bibinfo{journal}{Pisma Zh. Eksp. Teor. Fiz.}
  \textbf{\bibinfo{volume}{80}}, \bibinfo{pages}{389} (\bibinfo{year}{2004}).

\bibitem[{\citenamefont{Moore and Read}(1991)}]{MR}
\bibinfo{author}{\bibfnamefont{G.}~\bibnamefont{Moore}} \bibnamefont{and}
  \bibinfo{author}{\bibfnamefont{N.}~\bibnamefont{Read}},
  \bibinfo{journal}{Nucl. Phys. B} \textbf{\bibinfo{volume}{360}},
  \bibinfo{pages}{362} (\bibinfo{year}{1991}).

\bibitem[{\citenamefont{Volovik}(1992)}]{VolovikBook}
\bibinfo{author}{\bibfnamefont{G.~E.} \bibnamefont{Volovik}},
  \emph{\bibinfo{title}{Exotic properties of superfluid $^3$He}}
  (\bibinfo{publisher}{World Scientific}, \bibinfo{address}{Singapore},
  \bibinfo{year}{1992}).

\bibitem[{\citenamefont{Read and Green}(2000)}]{Read2000}
\bibinfo{author}{\bibfnamefont{N.}~\bibnamefont{Read}} \bibnamefont{and}
  \bibinfo{author}{\bibfnamefont{D.}~\bibnamefont{Green}},
  \bibinfo{journal}{Phys. Rev. B} \textbf{\bibinfo{volume}{61}},
  \bibinfo{pages}{10267} (\bibinfo{year}{2000}).

\bibitem[{\citenamefont{Landau and Lifshitz}(1981)}]{LL}
\bibinfo{author}{\bibfnamefont{L.~D.} \bibnamefont{Landau}} \bibnamefont{and}
  \bibinfo{author}{\bibfnamefont{E.~M.} \bibnamefont{Lifshitz}},
  \emph{\bibinfo{title}{Quantum Mechanics}}
  (\bibinfo{publisher}{Butterworth-Heinemann}, \bibinfo{address}{Oxford, UK},
  \bibinfo{year}{1981}).

\bibitem[{swa()}]{swave}
\bibinfo{note}{Contrast this with the $s$-wave BCS-BEC superfluid, where the
  dimensionless coupling is $m^{3 \over 2} g^2/\sqrt{\epsilon_F}$, that
  descreases at (not easily reachable) high density.}

\bibitem[{BCS()}]{BCScutoff}
\bibinfo{note}{This mean-field approximation remains valid even in the regime
  $\epsilon_\Lambda\gg \epsilon_F$ of interest to us, a limit that has not been
  considered in condensed matter systems.}

\bibitem[{\citenamefont{Vollhardt and W\"olfle}(1990)}]{Vollhardt}
\bibinfo{author}{\bibfnamefont{D.}~\bibnamefont{Vollhardt}} \bibnamefont{and}
  \bibinfo{author}{\bibfnamefont{P.}~\bibnamefont{W\"olfle}},
  \emph{\bibinfo{title}{The Superfluid Phases of He$^3$}}
  (\bibinfo{publisher}{Taylor \& Francis}, \bibinfo{address}{N.Y.},
  \bibinfo{year}{1990}).

\bibitem[{\citenamefont{Anderson and Morel}(1961)}]{Anderson1961}
\bibinfo{author}{\bibfnamefont{P.~W.} \bibnamefont{Anderson}} \bibnamefont{and}
  \bibinfo{author}{\bibfnamefont{P.}~\bibnamefont{Morel}},
  \bibinfo{journal}{Phys. Rev.} \textbf{\bibinfo{volume}{123}},
  \bibinfo{pages}{1911} (\bibinfo{year}{1961}).

\bibitem[{\citenamefont{Kitaev}(2003)}]{Kitaev2003}
\bibinfo{author}{\bibfnamefont{A.}~\bibnamefont{Kitaev}},
  \bibinfo{journal}{Ann. Phys.} \textbf{\bibinfo{volume}{303}},
  \bibinfo{pages}{2} (\bibinfo{year}{2003}).

\bibitem[{\citenamefont{Cheng and Yip}()}]{Yip2005}
\bibinfo{author}{\bibfnamefont{C.-H.} \bibnamefont{Cheng}} \bibnamefont{and}
  \bibinfo{author}{\bibfnamefont{S.-K.} \bibnamefont{Yip}},
  \eprint{cond-mat/0504278}.

\bibitem[{\citenamefont{Gurarie and Radzihovsky}(2005)}]{Gurarie2005}
\bibinfo{author}{\bibfnamefont{V.}~\bibnamefont{Gurarie}} \bibnamefont{and}
  \bibinfo{author}{\bibfnamefont{L.}~\bibnamefont{Radzihovsky}},
  \bibinfo{journal}{in preparation}  (\bibinfo{year}{2005}).

\end{thebibliography}

\end{document}